\documentclass[showpacs,twocolumn,nofootinbib,superscriptaddress,10pt,a4paper,pra]{revtex4-1}

\usepackage{natbib}
\usepackage{amsmath}
\usepackage{amssymb}
\usepackage{dsfont}
\usepackage{graphicx}
\usepackage{color}

\newcommand{\im}{{\rm i}}

\newcommand{\ket}[1]{|{#1}\rangle}
\newcommand{\bra}[1]{\langle{#1}|}
\newcommand{\braket}[2]{\langle{#1}|{#2}\rangle}

\begin{document}

\title{Turbulence and the Hong-Ou-Mandel effect}

\author{Shashi \surname{Prabhakar}}
\affiliation{School of Physics, University of the Witwatersrand, Johannesburg 2000, South Africa}
\affiliation{CSIR National Laser Centre, PO Box 395, Pretoria 0001, South Africa}

\author{Chemist \surname{Mabena}}
\affiliation{School of Physics, University of the Witwatersrand, Johannesburg 2000, South Africa}
\affiliation{CSIR National Laser Centre, PO Box 395, Pretoria 0001, South Africa}

\author{Thomas \surname{Konrad}}
\affiliation{University of Kwazulu-Natal, Private Bag X54001, Durban 4000, South Africa}
\affiliation{National Institute of Theoretical Physics, Durban Node, South Africa}

\author{Filippus S. \surname{Roux}}
\email{froux@nmisa.org}
\affiliation{National Metrology Institute of South Africa, Meiring Naud{\'e} Road, Brummeria, Pretoria, South Africa}
\affiliation{School of Physics, University of the Witwatersrand, Johannesburg 2000, South Africa}

\begin{abstract}
The effect of a decoherence channel, such as a turbulent atmosphere, on the second order quantum interference in the Hong-Ou-Mandel (HOM) effect is investigated. The investigation includes both theoretical analyses and an experimental implementation of the process. In our experiment, entangled input states are prepared with spontaneous parametric down-conversion. The atmospheric turbulence is modelled as a single-phase screen and simulated with a spatial light modulator according to the theory of Kolmogorov. We find both theoretically and experimentally that the HOM dip is unaffected when only one of the photons passes through turbulence, but both photons pass through turbulence, the HOM interference is only slightly affected by the scintillation. The reasons behind these findings and their consequences for HOM-based teleportation are discussed.
\end{abstract}

\maketitle

\section{Introduction}

Long distance quantum communication requires quantum repeaters that are based on quantum teleportation \cite{telezeil,repeater}. To implement quantum teleportation, one needs to perform joint measurements (Bell measurements), projecting a pair of photons onto a Bell state. One way to perform such a joint measurement is with the aid of the Hong-Ou-Mandel (HOM) effect \cite{hongoumandel}, according to which a photon pair entering both input ports of a balanced beamsplitter in an (anti)symmetric state leads to photon (anti)bunching in the output ports. A filter for antisymmetric Bell states can thus be realized by conditioning on coincidence photo-detection in the output ports of a balanced beamsplitter. The HOM effect can also be used as part of a protocol to synchronize clocks remotely with a high accuracy \cite{homsync1,homsync2}.

The teleportation process can be generalized to higher dimensional states \cite{teleoam,scissors} and has been applied to orbital angular momentum (OAM) modes \cite{nagali,pires,oamhom}. In long distance free-space quantum communication systems, it is expected that turbulence would affect the teleportation process and also the HOM effect, which is used to mediate the teleportation process. This expectation is based on the fact that turbulence can cause considerable distortion of optical modes traversing atmospheric channels, and thus may affect the quantum interference observed in the HOM effect. If the HOM synchronization protocol is implemented over an atmospheric channel \cite{sync1}, turbulence may in a similar way cause the synchronization process to fail.

In this paper, we perform a theoretical investigation of the effect of turbulence on the HOM interference effect and provide experimental confirmation for some of the results. It is shown that under certain conditions, turbulence does not have any effect on the quantum interference in the HOM effect. This opens up the use for quantum communication and quantum synchronization via free-space channels.

Two theoretical analyses are provided. One considers a one-sided turbulence channel, where one of the two photons after the SPDC process passes through turbulence under weak scintillation conditions, while the other photon does not experience any turbulence. The other analysis investigates a two-sided turbulence channel, where both photons after the SPDC process pass through turbulence under weak scintillation conditions. The former analysis is simpler, because one can exploit the orthogonality of the modal basis to simplify the analysis. It clearly demonstrates that the HOM interference effect is not affected by one-sided weak scintillation. The latter analysis is more involved and requires a more thorough calculation. It shows that, although not completely unaffected, the HOM interference effect is only slightly modified by the scintillation in a two-sided turbulence channel.

In the experiment, the input state of a photon pair is prepared with spontaneous parametric down-conversion (SPDC). The turbulence is simulated by random phase modulations with the aid of phase-only spatial light modulators (SLMs). As such, the turbulence is represented by a single-phase screen \cite{paterson}, which is valid for weak scintillation conditions \cite{turbsim}. The random phase modulations on the SLMs are computed according to the Kolmogorov theory \cite{scintbook} in the experiment. However, for tractability, a quadratic structure function approximation \cite{leader} is used in the calculations. The HOM interference is obtained with the aid of a balanced (50:50) beamsplitter --- the HOM filter. The SPDC process naturally produces a symmetric state, which leads to a lack of coincidence counts (a HOM dip) after the HOM filter. We measure the influence of turbulence on the visibility of the HOM dip. The photons after the HOM filter are observed in the Laguerre-Gauss (LG) basis, using additional SLMs behind the two output ports.

The paper is organized as follows. We start in Sec.~\ref{oneside} by discussing the situation for a one-sided channel and in Sec.~\ref{twoside} we discuss the situation for a two-sided channel. The experimental setup is explained and the results are presented in Sec.~\ref{expsetup}. We provide some discussion in Sec.~\ref{disc} and give a final summary in Sec.~\ref{summ}.

\section{One-sided weak scintillation}
\label{oneside}

First, we consider the special situation where only one photon passes through turbulence and where the turbulence conditions are such that it only causes a weak scintillation. Both in the analytical calculations and in the experimental observations for this case, one finds the counter-intuitive result that the HOM dip remains equally deep regardless of the strength of the scintillation. Although the coincidence counts decrease with increasing scintillation strength, the visibility of the HOM dip remains constant (apart from fluctuations, caused by an increase in the shot-noise). The relative depth of the dip (ratio of counts inside and outside the dip) or its visibility (counts outside minus counts inside divided by counts outside plus counts inside) remains the same. Intuitively, one would have expected the scintillation to pollute the initial symmetric input state with an antisymmetric component. Since the latter adds coincidence counts inside the dip, it should gradually fill in the dip. However, this does not happen. Here, we explore the physics behind this phenomenon and clarify the conditions for this particular robustness.

The experimental conditions (see more detail below) for observation of the robustness of the HOM dip are as follows: (a) The input state is an SPDC state, which is symmetric with respect to an exchange of the photon-paths. (b) The turbulence is modelled as a single-phase screen (assuming weak scintillation conditions). (c) The single-phase screen for the turbulence is only placed in one of the two paths, representing a one-sided turbulence channel. (d) The output after the HOM filter (beamsplitter) is measured by means of projections onto particular LG modes in the respective output ports. These modes all have radial index $p=0$ and the magnitude of their azimuthal index is $|\ell|=1$.

The state obtained from the SPDC process can be expressed in terms of an expansion of symmetric Bell-states $\ket{\Psi_{\ell}^{+}}$, consisting of LG modes with the same magnitude of the azimuthal index $\ell$. The symmetric (antisymmetric) Bell-state $\ket{\Psi_{\ell}^{+}}$ ($\ket{\Psi_{\ell}^{-}}$) is defined by
\begin{equation}
\ket{\Psi^{\pm}_{\ell}} = \frac{1}{\sqrt{2}}\left(\ket{\ell}_A\ket{\bar{\ell}}_B\pm\ket{\bar{\ell}}_A\ket{\ell}_B\right) ,
\label{defpsi}
\end{equation}
where $\bar{\ell}=-\ell$. The SPDC state can thus be regarded as a linear superposition of the symmetric Bell states
\begin{equation}
\ket{\psi_{\rm SPDC}} = \sum_{\ell=0}^{\infty} \ket{\Psi^{+}_{\ell}} \alpha_{\ell} ,
\label{defspdc}
\end{equation}
where $\alpha_{\ell}$ denotes the expansion coefficients.

The reason for the robustness of the HOM dip must lie in the probability for the one-sided turbulence channel to convert a symmetric Bell state into an antisymmetric Bell state (or vice versa). To see this, we consider the transition amplitude
\begin{equation}
\eta_{\ell} = \bra{\Psi_{\ell}^{-}}\mathds{1}_A\otimes \hat{T}_B\ket{\psi_{\rm SPDC}} ,
\end{equation}
where $\hat{T}_B$ is the turbulence operator (in path $B$) representing a particular realization of the turbulence;\footnote{In the statistical analysis, the results are averaged over all possible realizations. However, to observe on average a vanishing coincidence count for the HOM dip, all the individual realizations must show vanishing transition amplitudes $\eta_\ell=0$ in the dip. Therefore, we need to consider the individual realizations.} and $\mathds{1}_A$ is the identity operator (in path $A$). The robustness of the HOM dip implies that $\eta_{\ell}=0$ for all $\ell$, irrespective of the scintillation strength.

By substituting the expanded SPDC state, given in Eq.~(\ref{defspdc}), into the transition amplitude, we obtain
\begin{equation}
\eta_{\ell} = \sum_{m=0}^{\infty} \bra{\Psi^{-}_{\ell}}\mathds{1}_A\otimes \hat{T}_B\ket{\Psi^{+}_{m}} \alpha_{m} .
\end{equation}
Due to the identity operation in path $A$, initial and final azimuthal indices must coincide: $m=\ell$. Hence, the only term that can contribute to the transition is
\begin{equation}
\eta_{\ell} = \bra{\Psi^{-}_{\ell}}\mathds{1}_A\otimes \hat{T}_B\ket{\Psi^{+}_{\ell}} \alpha_{\ell} .
\label{tramp}
\end{equation}
Thus we see that $\eta_{\ell}$ reduces to the transition amplitude between the symmetric Bell state and its corresponding antisymmetric Bell state.

Next, we substitute Eq.~(\ref{defpsi}) into Eq.~(\ref{tramp}) and evaluate the inner products in path $A$ (the side without turbulence), to obtain
\begin{eqnarray}
\eta_{\ell} & = & \frac{\alpha_{\ell}}{2}\left(\bra{\bar{\ell}}_B \hat{T}_B \ket{\bar{\ell}}_B -\bra{\ell}_B \hat{T}_B \ket{\ell}_B \right) \nonumber \\
& = & - \frac{\alpha_{\ell}}{2} {\rm tr}\{ \hat{T}_B \hat{\sigma}_z \} ,
\end{eqnarray}
where $\hat{\sigma}_z$ is the operator for the Pauli $z$-matrix, defined in the $|\ell|$-subspace. It thus turns out that the transition amplitude is proportional to the coefficient of the $\hat{\sigma}_z$ component of the turbulence operator, when expanded as a Bloch representation for a qubit channel.

To determine the coefficient for $\hat{\sigma}_z$ in the Bloch representation of $\hat{T}$, we convert the expression to the spatial domain. For this purpose, we define the single-phase screen turbulence operator by
\begin{equation}
\hat{T} = \int \ket{{\bf x}} \exp[\im\theta_i({\bf x})] \bra{{\bf x}}\ {\rm d}^2x ,
\end{equation}
where $\theta_i({\bf x})$ is a particular realization of a random phase function that represents the effect of the turbulence. Applying this definition in the expression for the transition amplitude, we obtain
\begin{eqnarray}
\eta_{\ell} & = & \frac{\alpha_{\ell}}{2} \int \braket{\bar{\ell}}{{\bf x}} \exp[\im\theta_i({\bf x})] \braket{{\bf x}}{\bar{\ell}} \nonumber \\
& & -\braket{\ell}{{\bf x}} \exp[\im\theta_i({\bf x})] \braket{{\bf x}}{\ell}\ {\rm d}^2x .
\label{etadef}
\end{eqnarray}

Next, we define the OAM modes in polar coordinates
\begin{equation}
\braket{{\bf x}}{\ell} = R_{|\ell|}(r)\exp(\im\ell\phi) ,
\end{equation}
and apply them in the expression. Since the turbulence is represented by a single local phase modulation (all located in the same plane), the input and output phase functions of the modes cancel each other, leaving only the $r$-dependent parts of the modes, which only depend on the magnitudes of the azimuthal index. These functions are the same for modes with the same magnitude of the azimuthal index. As a result, the two terms in Eq.~(\ref{etadef}) are equal and cancel each other, leaving us with $\eta_{\ell}=0$. So we see that the transition amplitude is zero and the Bloch representation of $\hat{T}$ does not contain $\hat{\sigma}_z$.

This explains why the HOM dip remains well-defined, regardless of the strength of the scintillation. However, it is important to note that this explanation relies on the fact that the turbulence is simulated with a single-phase screen and that it only acts on one of the two photons. Under different conditions this mechanism may not apply anymore.

\section{Two-sided weak scintillation}
\label{twoside}

Next, we consider the situation where both photons pass through turbulence. The turbulence conditions are still those that correspond to weak scintillation. The analytical calculations and the experimental observations for this case reveal a slight decrease in the quality of the HOM dip with increasing strength of the scintillation. The experimental conditions for the two-sided case are the same as for the one-sided case, stated at the beginning of Sec.~\ref{oneside}, with the exception that single-phase screens, simulating turbulence, are placed in both paths.

The theoretical calculations for the two-sided case are more involved, because two independent azimuthal indices have to be taken into account. As a result, we need to consider the calculations for this case in more detail.

Generically, one can represent the biphoton input state, as produced by the SPDC process, in terms of its spatial degrees of freedom by
\begin{equation}
\ket{\psi} = \int \ket{{\bf R}_1}_A \ket{{\bf R}_2}_B \psi({\bf R}_1,{\bf R}_2)\ {\rm d}^3r_1\ {\rm d}^3r_2 ,
\label{input}
\end{equation}
where $\ket{{\bf R}}$ represents a three-dimensional coordinate basis. In other words, we include the $z$-dependence, which relates to the path length, in addition to the two transverse ($x,y$) dependences. The two photons are separated into different paths (or channels), labeled $A$ and $B$, respectively. The HOM dip is observed as a function of the relative path lengths, centered at zero relative path length.

To consider the effect of the path length difference for an SPDC state, we convert the expression to the Fourier domain in the $z$-dependence. Hence, we have
\begin{eqnarray}
\ket{\psi} & = & \int \ket{{\bf r}_1,c_1}_A \ket{{\bf r}_2,c_2}_B F({\bf r}_1,{\bf r}_2) \nonumber \\
& & \times \exp(\im 2\pi c_1 dz) \exp(-\im 2\pi c_2 dz) h(2 c_0-c_1-c_2) \nonumber \\
& & \times H(c_1) H(c_2)\ {\rm d}^2r_1\ {\rm d}^2r_2\ {\rm d}c_1\ {\rm d}c_2 ,
\label{input0}
\end{eqnarray}
where ${\bf r}_1$ and ${\bf r}_2$ are two-dimensional transverse position vectors; $c_1$ and $c_2$ are the longitudinal spatial frequency components (directly related to the wavelength of the light) associated with the $z$-components of the position vectors. The spatial correlations (due to momentum conservation in the SPDC process) are incorporated into $F({\bf r}_1,{\bf r}_2)$ and the wavelength correlation (due to energy conservation in the SPDC process) is captured by $h(2 c_0-c_1-c_2)$, where $c_0$ is the longitudinal spatial frequency that is associated with the wavelength of the pump beam. The nature of $h(\cdot)$ is discussed below. The $z$-dependences for the two photons are refined so that the nominal path length is set to zero, leaving only the relative path length $dz$ with opposite signs for the two photons. The individual spectra of the two photons, as determined by wavelength filters, are given by $H(c_1)$ and $H(c_2)$, respectively (see detail below).

Both photons of the input state pass through simulated turbulence. We use unitary operators $\hat{U}_A$ and $\hat{U}_B$ for the respective channels, to represent the scintillation process caused by the turbulence. For single-phase screen turbulence (weak scintillation), these unitary operators implement phase modulations $\hat{U} \ket{{\bf r}}=\ket{{\bf r}}\exp[\im\theta({\bf r})]$. Hence
\begin{eqnarray}
\hat{U}_A \ket{{\bf r}_1,c_1}_A & = & \ket{{\bf r}_1,c_1}_A \exp[\im\theta_A({\bf r}_1)] \nonumber \\
\hat{U}_B \ket{{\bf r}_2,c_2}_B & = & \ket{{\bf r}_2,c_2}_B \exp[\im\theta_B({\bf r}_2)] .
\label{turb}
\end{eqnarray}

\begin{figure}[ht]
\includegraphics{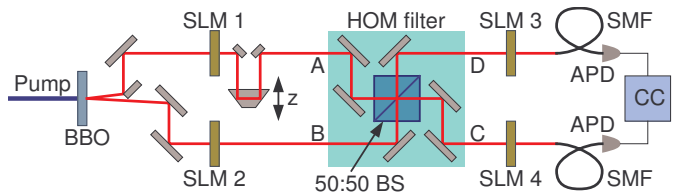}
\caption{Experimental setup used to observe the Hong-Ou-Mandel dip after the quantum state has passed through one-sided or two-sided single-phase screen turbulence.}
\label{opstel}
\end{figure}

After the turbulence, the photons in paths $A$ and $B$ are, respectively, sent into the two input ports of a balanced beamsplitter, as shown in Fig.~\ref{opstel}, to perform the HOM filtering and obtain HOM interference. The unitary operator $\hat{U}_{\rm bs}$ that represents this beamsplitter, performs the following transformations
\begin{eqnarray}
\ket{{\bf r}_1,c_1}_A & \rightarrow & \frac{1}{\sqrt{2}} \left( \ket{{\bf r}_1,c_1}_C + \ket{{\bf r}_1,c_1}_D \right) \nonumber \\
\ket{{\bf r}_2,c_2}_B & \rightarrow & \frac{1}{\sqrt{2}} \left( \ket{{\bf r}_2,c_2}_C - \ket{{\bf r}_2,c_2}_D \right) ,
\label{bsdef}
\end{eqnarray}
where the output ports are denoted by $C$ and $D$, as shown in Fig.~\ref{opstel}. Mirrors are added in the setup for the HOM filter so that every path from $A$ or $B$ to $C$ or $D$ have an even number of reflections, thus avoiding the inversion of the helicity of the modes. After the beamsplitter operation, we condition on observing one photon in coincidence in each of the two output ports. The result, just before detection, reads,
\begin{equation}
\ket{\psi'} = \hat{\cal P}_{\rm CC} \hat{U}_{\rm bs} \hat{U}_A \hat{U}_B \ket{\psi} ,
\label{uitset1}
\end{equation}
where $\hat{\cal P}_{\rm CC}$ is a projection operator that selects out the coincidence part of the state.

To observe the HOM dip, we use SLMs to project the respective photon states onto particular spatial modes. For this purpose, we select a pair of LG modes with opposite azimuthal indices for the two respective photons, under condition of coincidence detection. The longitudinal degrees of freedom are not affected (filtered) in this process. The result is a post-selection in the transverse degrees of freedom onto a two-dimensional subspace of LG modes. We express the result of this projective measurement as the detection probability $P$, given by
\begin{equation}
P = {\rm tr}\{\hat{\cal P}_{\ell} \hat{\rho}\} = \bra{\psi'} \hat{\cal P}_{\ell} \ket{\psi'} ,
\label{meet}
\end{equation}
where $\hat{\rho}=\ket{\psi'}\bra{\psi'}$ is the density operator for the state after the beamsplitter and $\ket{\psi'}$ is given in Eq.~(\ref{uitset1}). The projection operator, associated with this projective measurement, is given by
\begin{equation}
\hat{\cal P}_{\ell} = \mathds{1}_z\otimes(\ket{\ell_C,\bar{\ell}_D}\bra{\ell_C,\bar{\ell}_D}) ,
\label{meetpro}
\end{equation}
where $\mathds{1}_z$ is an identity operator for the longitudinal degrees of freedom (associated with $z$). By implication, we assume that the LG modes are independent of the wavelength
\begin{equation}
\braket{{\bf r},c_1}{\ell} = \braket{{\bf r}}{\ell} = u_{\ell}({\bf r}) ,
\label{rho1}
\end{equation}
where $u_{\ell}({\bf r})$ is the LG mode with azimuthal index $\ell$.

Note that, because the longitudinal degrees of freedom are not filtered, the expression for $P$ would contain the squares of the functions that depend on $c_1$ and $c_2$. In the measurement process, the longitudinal degrees of freedom are traced over, which means that we need to integrate over $c_1$ and $c_2$. For this purpose, we replace the square of the $h$-function with a Dirac delta function\footnote{The function $h(\cdot)$ imposes energy conservation. However, if we assume that it is given by a Dirac delta function $h(\cdot)=\delta(\cdot)$, then we will end up with a squared Dirac delta function, which would give a divergent result.}
\begin{equation}
h^2(2 c_0-c_1-c_2) \rightarrow \delta(2 c_0-c_1-c_2) .
\label{rho}
\end{equation}
We redefine the integration variables $c_1=c_0+c_3$ and $c_2=c_0+c_4$. The Dirac delta function will then assign $c_4=-c_3$ after integrating over $c_4$. We'll assume that the spectra can be defined in terms of a hat-function
\begin{equation}
H(c) = \Pi\left(\frac{c-c_c}{W}\right) ,
\label{spekdef}
\end{equation}
where $W$ is the width of the spectrum (determined by the bandwidth of the line filters), $c_c$ is the center frequency (of the line filters) and
\begin{equation}
\Pi(x) = \left\{ \begin{array}{cc} 1 & {\rm for}~|x|<1/2 \\ 0 & {\rm otherwise} \end{array} \right. .
\label{pidef}
\end{equation}
Note that $\Pi^2(x)=\Pi(x)$. In the end, we have terms that contain either of two possible integrals over $c_3$:
\begin{eqnarray}
\int H^2(c_0+c_3) H^2(c_0-c_3)\ {\rm d}c_3 & = & W , \nonumber \\
\int H^2(c_0+c_3) H^2(c_0-c_3) & & \nonumber \\
 \times \exp(\pm \im 8\pi c_3 dz)\ {\rm d}c_3 & = & W {\rm sinc}(4\pi W dz) . \nonumber \\
\label{hint2}
\end{eqnarray}

The random phase functions $\theta({\bf r})$ in Eq.~(\ref{turb}) represent specific realizations of the turbulent medium. We compute the ensemble average over all such realizations in both photon paths, independently, to obtain the averaged coincidence counts. The effect of the ensemble averaging is to convert the random phase factors into exponential functions
\begin{equation}
{\cal E}\{\exp[\im\theta({\bf r}_1)-\im\theta({\bf r}_2)]\} = \exp\left[-\frac{1}{2}D({\bf r}_1-{\bf r}_2)\right] ,
\label{struktf}
\end{equation}
that contain the phase structure function $D({\bf r}_1-{\bf r}_2)$ \cite{scintbook}. One can now express the ensemble averaged detection probability by
\begin{eqnarray}
P & = & \int \exp\left[-\frac{1}{2}D({\bf r}_1-{\bf r}_3)\right] \exp\left[-\frac{1}{2}D({\bf r}_2-{\bf r}_4)\right] \nonumber \\
 & & \times \left[F({\bf r}_1,{\bf r}_2) F^*({\bf r}_3,{\bf r}_4)+F({\bf r}_2,{\bf r}_1) F^*({\bf r}_4,{\bf r}_3) \right] \nonumber \\
 & & \times \left[ u_{\ell}^*({\bf r}_1) u_{\bar{\ell}}^*({\bf r}_2) - {\rm sinc}(4\pi W dz) u_{\ell}^*({\bf r}_2) u_{\bar{\ell}}^*({\bf r}_1)\right] \nonumber \\
 & & \times u_{\ell}({\bf r}_3) u_{\bar{\ell}}({\bf r}_4)\ {\rm d}^2r_1\ {\rm d}^2r_2\ {\rm d}^2r_3\ {\rm d}^2r_4 .
\label{meet5}
\end{eqnarray}

For the purpose of the calculations, we'll assume $|\ell|=1$ and radial index $p=0$. In Cartesian coordinate, the LG modes with these indices are given by
\begin{equation}
u_{\pm 1}({\bf r}) = (x\pm\im y) \exp\left(-\frac{|{\bf r}|^2}{w_0^2}\right) ,
\label{lgdef}
\end{equation}
where $w_0$ is the mode size for the LG mode and where we ignore the normalization constant.

The function $F({\bf r}_1,{\bf r}_2)$, which represents the spatial correlation of the down-converted state, can be represented in the Fourier domain as the product of the pump beam and the phase matching function. The argument of the pump beam consists of the sum of the Fourier domain coordinate vectors of the down-converted beams, due to the transverse phase matching condition (momentum conservation). The argument of the phase matching function, on the other hand, is the difference of the Fourier domain coordinate vectors of the down-converted beams, as a result of the paraxial condition and the phase matching condition.

Often, the phase matching function is approximated by a Gaussian function \cite{law}, which gives more tractable expressions than the sinc-function found in the expression of the actual phase matching function. Such an approximation is only valid in the thin crystal limit \cite{pindex}. Beyond this limit, the sub-leading term of the Gaussian approximation gives a different scale behavior than that of the sinc-function
\begin{eqnarray}
\exp(-\beta |{\bf k}|^2) & \approx & 1 - \beta |{\bf k}|^2 + {\rm O}\left(\beta^2\right) \\
{\rm sinc}(\beta |{\bf k}|^2) & \approx & 1 - \frac{1}{6} \beta^2 |{\bf k}|^4 + {\rm O}\left(\beta^4\right) .
\end{eqnarray}
As a result, these two functions would give inconsistent predictions for the behavior beyond the thin crystal limit (for $\beta\neq 0$). We are interested in the behavior beyond the thin crystal limit. Therefore, we need to consider the phase matching function in terms of the sinc-function. To alleviate the calculation, we'll use an auxiliary integral to represent the sinc-function \cite{noncol}
\begin{equation}
{\rm sinc}(x) = \frac{1}{2} \int_{-1}^1 \exp(\im x\xi)\ {\rm d}\xi .
\label{auxsinc}
\end{equation}

After an inverse Fourier transform, the spatial correlation function is expressed as
\begin{eqnarray}
F({\bf r}_1,{\bf r}_2) & = & \frac{\cal N}{2 w_p^2} \exp\left( -\frac{|{\bf r}_1+{\bf r}_2|^2}{4 w_p^2} \right)\nonumber \\
 & & \times \int_{-1}^1 \frac{1}{\xi} \exp\left( -\im \frac{|{\bf r}_1-{\bf r}_2|^2}{2\xi\beta w_p^2} \right)\ {\rm d}\xi ,
\label{spakor}
\end{eqnarray}
where ${\cal N}$ is a normalization constant, $w_p$ is the pump beam radius and $\beta$ is the crystal ratio, defined as the crystal length $L$ (times the ordinary refractive index of the crystal $n_o$) divided by the pump Rayleigh range $z_R$
\begin{equation}
\beta = \frac{n_o L}{z_R}= \frac{n_o L\lambda}{\pi w_p^2} ,
\label{betadef}
\end{equation}
with $\lambda$ being the pump wavelength. The thin crystal limit is obtained by taking the limit $\beta\rightarrow 0$.

To enable the analytic evaluation of the integrals, we use a quadratic structure function approximation \cite{leader},
\begin{equation}
D(\Delta {\bf r}) = 6.88 {\cal W}^{5/3}\frac{|\Delta {\bf r}|^2}{w_p^2} ,
\label{qsfdef}
\end{equation}
where
\begin{equation}
{\cal W} = \frac{w_p}{r_0} ,
\label{wdef}
\end{equation}
with
\begin{equation}
r_0 = 0.185 \left(\frac{\lambda^2}{C_n^2 z}\right)^{3/5} ,
\label{fried}
\end{equation}
being the Fried parameter \cite{fried}. Here, $C_n^2$ is the refractive index structure constant, which quantifies the strength of the turbulence, and $z$ is the propagation distance.

Substituting Eqs.~(\ref{lgdef}), (\ref{spakor}) and (\ref{qsfdef}) into Eq.~(\ref{meet5}), one can evaluate the integrals. Those that involve ${\bf r}_1$, ${\bf r}_2$, ${\bf r}_3$ and ${\bf r}_4$ are readily evaluated. The remaining auxiliary integrations (associated with the sinc-functions) are somewhat more challenging, but still tractable. The resulting expression for the probability to measure coincidence counts is rather complicated. Therefore, we do not provide the detailed expression. However, we find that the expression has the form
\begin{eqnarray}
P & = & f_1(\alpha,\beta,\zeta) \left[1-{\rm sinc}(4\pi W dz)\right] \nonumber \\
 & & + \zeta^2\beta^4 f_2(\alpha,\beta,\zeta) ,
\label{uitset2}
\end{eqnarray}
where $\beta$ is defined in Eq.~(\ref{betadef}) and
\begin{eqnarray}
\alpha & = & \frac{w_0^2}{w_p^2} \label{alphadef} \\
\zeta & = & 6.88 {\cal W}^{5/3} . \label{zetadef}
\end{eqnarray}
The factor $[1-{\rm sinc}(4\pi W dz)]$ represents the shape of a perfect HOM dip. The first term in Eq.~(\ref{uitset2}) by itself gives a constant dip, while $f_1(\alpha,\beta,\zeta)$ denotes the level of coincidence counts for the curve. The second term, with $f_2(\alpha,\beta,\zeta)$ serves to fill in the dip and thereby reduces its quality. However, we see that the second term contains a factor of $\zeta^2$, which means that, for no turbulence, it is zero. We also see that it contains a factor of $\beta^4$, which implies that this term is quite small, because the value of $\beta$ tends to be small in typical experiments.\footnote{Under the Gaussian approximation of the phase matching function, the second term would contain $\beta^2$ instead of $\beta^4$ and would thus overestimate the effect of turbulence on the quality of the HOM dip.} In the thin crystal limit, where $\beta\rightarrow 0$, the second term is zero. Therefore, we see that, even for the two-sided channel, the effect of the turbulence on the quality of the dip is quite small and would often be negligible. In fact, the second term, which would fill in the dip, depends on experimental parameters (the crystal length and the pump mode size) which can be chosen in the experiment as to minimized its effect.

\begin{figure*}[ht]
\includegraphics{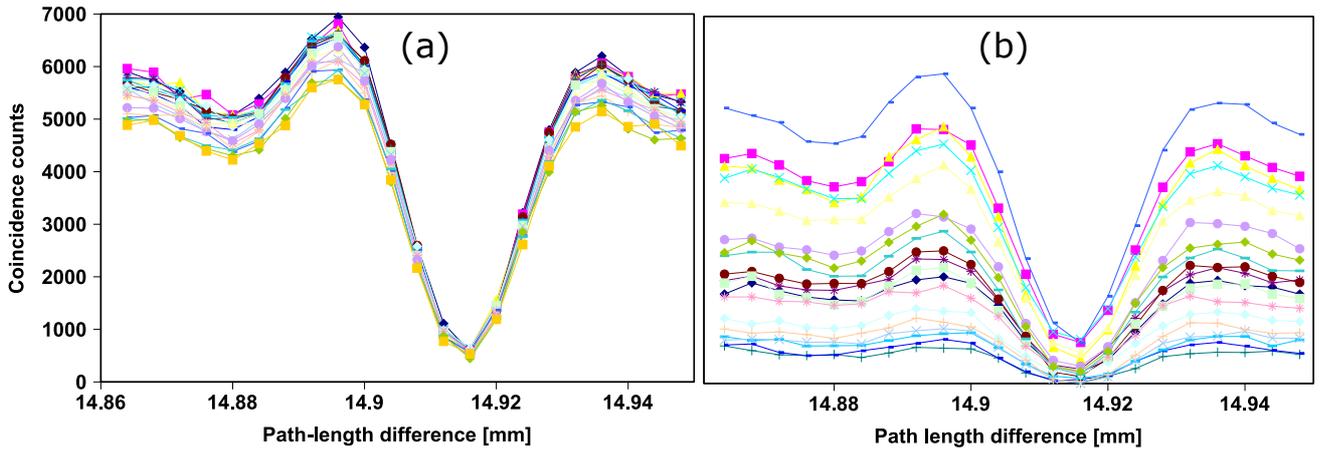}
\caption{Experimentally observed HOM dip curves in the one-sided case for (a) ${\cal W}=0$ and (b) ${\cal W}=0.4$.}
\label{rou}
\end{figure*}

\section{Experimental setup and results}
\label{expsetup}

The experimental setup is shown in Fig.~\ref{opstel}. A mode-locked laser source with a wavelength of 355~nm, an average power of 350~mW and a repetition rate of 80~MHz is used. The pump beam has a beam radius of 58.9~$\mu$m. It pumps a 3~mm-thick type I BBO crystal to produce noncollinear, degenerate photon pairs via SPDC. A small noncollinear angle of $\sim$3 degrees exists between the signal and idler beams. The plane of the crystal is imaged onto SLM1 and SLM2, via paths $A$ and $B$, respectively, with a magnification of $\times 4$ (4-f system with $f_{1} = 100$~mm and $f_{2} = 400$~mm not shown). The atmospheric turbulence is simulated by means of random phase modulations, using either SLM1 (in path $A$) for the single-sided case or with both SLM1 and SLM2 (in path $A$ and $B$, respectively) for the double-sided case.

\begin{figure}[ht]
\includegraphics{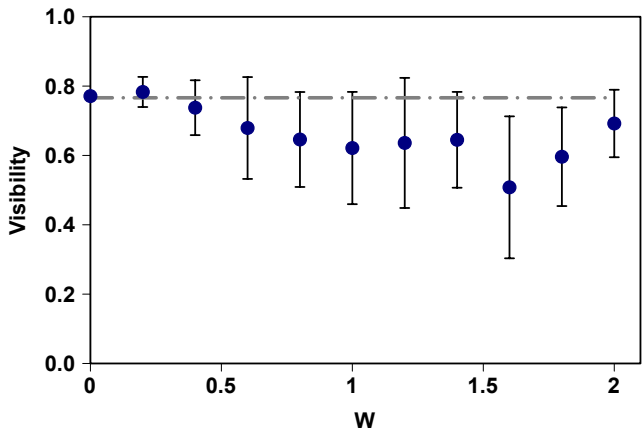}
\caption{Visibility of the HOM dip as a function of ${\cal W}$ for the case of a one-sided turbulent channel. Experimental values are shown as blue circles with vertical error bars, denoting the standard deviation. The horizontal gray dashed line represents the best fit constant visibility.}
\label{sssym}
\end{figure}

The two-mirror combination in path $A$, which is used to change the relative path length between the two paths, is mounted on a motorized translation stage. This allows one to scan through the dip in coincidence counts as a function of the relative path length difference $dz$. The plane of SLM1 and SLM2 are imaged onto SLM3 and SLM4 without magnification (4f system with $f_{3} = 500$~mm and $f_{4} = 500$~mm not shown).

\begin{figure}[ht]
\includegraphics{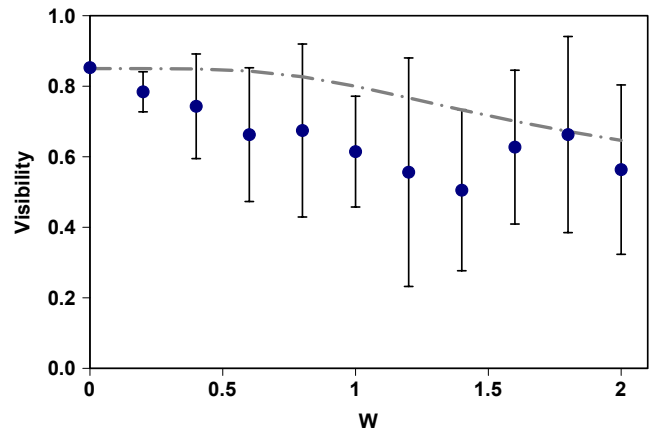}
\caption{Visibility of the HOM dip as a function of ${\cal W}$ for the case of a two-sided turbulent channel. Experimental values are shown as blue circles with vertical error bars, denoting the standard deviation. The gray dashed curve represents the theoretically calculated result, but with a lower initial visibility to fit the experimental value for no turbulence.}
\label{dssym}
\end{figure}

By selecting particular LG modes for detection, projective measurements are performed using SLM3 and SLM4, together with the single mode fibers (SMFs). We detect $\ell=1$ on SLM3 and $\ell=-1$ on SLM4. The mode size of the LG modes programmed on SLM3 and SLM4 is $w_0=450.0~\mu$m. The SLM3 and SLM4 planes are re-imaged with a demagnification factor of $\times 375$ (4-f system with $f_{5} = 750$~mm and $f_{6} = 2$~mm not shown) onto the SMFs. The back-propagated beams from the SMFs have beam radii of 882~$\mu$m and 838~$\mu$m on the planes of SML3 and SLM4, respectively. After SLM3 and SLM4, the two beams pass through 10~nm bandwidth interference filters (IF) before coupling into the SMFs. Avalanche photo diodes (APDs) at the ends of the SMFs are used to register the photon pairs with the aid of a coincidence counter (CC). The measured coincidence counts are accumulated over a 25~s integration time, with a gating time of 12.5~ns (based on the repetition rate of the laser).

The experimental parameters give $\alpha=4.19$ and $\beta=0.173$. They were specifically chosen to give a value for $\beta$ that is not too small.

We performed several experimental runs for each of the different values of the parameter ${\cal W}$, which ranged from 0 to 2. In Fig.~\ref{rou}, we show the experimentally observed curves of the HOM dip in the case of a one-sided turbulence channel. The curves in Fig.~\ref{rou}(a) are for no turbulence, ${\cal W}=0$. One can see well-defined curves with very little variation among the different runs. With a moderate scintillation of ${\cal W}=0.4$ in the one-sided turbulence channel, one observes much more fluctuation in the curves for the different runs, as seen in Fig.~\ref{rou}(b). Although there is a relatively large variation in the levels of coincidence counts among these curves, the shape of the dip is still clearly recognizable in them.

For each of the curves, we computed the visibility of the dip, defined as
\begin{equation}
{\cal V} = \frac{C_{\rm out}-C_{\rm in}}{C_{\rm out}+C_{\rm in}} ,
\label{visdef}
\end{equation}
where $C_{\rm out}$ and $C_{\rm in}$ are the coincidence counts outside and inside the dip, respectively. We computed the averages and standard deviations of these visibilities for each value of ${\cal W}$. The result for the one-sided case is shown in Fig.~\ref{sssym}. We also show a constant value as a gray dashed line, representing the best-fit value for the constant visibility of ${\cal V} = 0.77\pm 0.03$. Due to experimental imperfections, this value is smaller than the ideal value of 1. We see that the experimental results maintain this constant value with good agreement, even up to a value of ${\cal W}=2$, which is considered to be a fairly large value for ${\cal W}$; variations in the behavior is usually expected to occur in the vicinity of ${\cal W}=1$ for $\ell=1$ \cite{oamturb}.

The visibilities for the two-sided case are shown in Fig.~\ref{dssym}. Here, we also show the curve for the theoretical calculation as a gray dashed curve. The theoretical curve has been multiplied with a factor to lower the initial value down from 1 so that it would match the lower visibility of the experimental results when no turbulence is present. The theoretical curve remains at a fairly high value of the visibility, even up to a value of ${\cal W}=2$. The experimental results follow this trend fairly well.

\section{Discussion}
\label{disc}

The fact that the HOM dip remains intact after the photons have passed through turbulence, may seem as a fortuitous consequence of second order quantum interference. However, what our analysis reveals is that this phenomenon is more a result of the properties of turbulence than of second order quantum interference. The fact that turbulence does not convert symmetric states into antisymmetric states is responsible for the robustness of the HOM interference effect. The latter simply produces a dip when provided with a symmetric input state.

The analysis presented in Sec.~\ref{oneside} ignores the radial degree of freedom. In a more general case, the individual Bell states, which are defined in terms of LG modes, can also have an additional summation over all the different $p$ indices. If the LG modes are the Schmidt basis for the SPDC state, then these summations over the radial index would break up into individual Bell states in which the radial indices appear in the same way as the azimuthal indices. One would then be able to select a particular set of radial indices just like we selected a particular set of azimuthal indices and follow the argument through, as presented in Sec.~\ref{oneside}. As a result, the radial index would not affect the argument. Unfortunately, the LG states are not the exact Schmidt basis for the SPDC process \cite{miatto3}. However, one can argue that the actual Schmidt basis is not significantly different from the LG modes; under the Gaussian approximation of the phase-matching function \cite{law}, the Schmidt basis does turn out to be the LG modes \cite{miatto4}. As a result, we don't expect the radial index to play a significant role in the analysis of Sec.~\ref{oneside}.

The analyses that are reported here are restricted to the domain of weak scintillation. It may therefore seem to be an artifact of the single-phase screen approximation that bestows this property on turbulence that it does not convert symmetric states into antisymmetric states. Indeed, when stronger scintillation is considered we do expect to see a gradual increase in the conversion of symmetric states into antisymmetric states. However, we still expect this process to be small compared to other processes that convert one symmetric state into another symmetric state, for instance.

A possible source of difference between the theoretical results and the experimental results is the fact that the Kolmogorov structure function is used to produce the phase screens in the experiment, whereas the structure function is approximated by a quadratic function in the theoretical analysis for the two-sided case to make the integrals tractable. However, the consistency between the theoretical results and the experimental results, as shown in Fig.~\ref{dssym}, indicates that this difference does not have a significant effect on the results.

\section{Summary}
\label{summ}

We have shown, both theoretically and experimentally that the HOM interference is not affected by one-sided turbulence under weak scintillation conditions. The reason for this phenomenon is revealed to be due to the fact that, in weak scintillation condition, turbulence does not convert symmetric states into antisymmetric states. We have also shown that even for a two-sided channel, the effect of turbulence on the quality of the HOM dip remains small under weak scintillation conditions. Moreover, with an appropriate choice of the crystal length and the pump mode size, the effect of turbulence for a two-sided channel can be reduced to a negligible level.

As an implication, free-space implementation of long distance quantum communication, has a better chance to succeed than one might have thought. The same is true for remote clock synchronization protocols based on second order quantum interference.

\section*{Acknowledgements}

The research was done with the partial support of a grant from the National Research Foundation (NRF).


\end{document}